\documentclass[11pt]{article}
\usepackage{Blois,epsfig}

\def\Journal#1#2#3#4{{#1} {\bf #2}, #3 (#4)}

\def\NIMA{{\em Nucl. Instrum. Methods} A}

\def\PLB{{\em Phys. Lett.}  B}
\def\PRL{\em Phys. Rev. Lett.}

\def\be{\begin{equation}}
\def\ee{\end{equation}}
\def\bea{\begin{eqnarray}}
\def\eea{\end{eqnarray}}

\begin{document}
\vspace*{2cm}
\begin{center}
\Large{\textbf{XIth International Conference on\\ Elastic and Diffractive Scattering\\ Ch\^{a}teau de Blois, France, May 15 - 20, 2005}}
\end{center}

\vspace*{2cm}
\title{NEW DIFFRACTION RESULTS FROM CDF}

\author{ C. MESROPIAN}

\address{The Rockefeller University, 1230 York Avenue, \\
New York, NY 10021, USA}

\maketitle\abstracts{
We report measurements of  hard diffractive processes performed by the
CDF
collaboration in proton-antiproton collisions at the Fermilab
Tevatron collider at $\sqrt{s}$=1960 GeV. 
The characteristics of
the diffractive structure function from diffractive dijet production
studies are presented. The results of
exclusive dijet production in double pomeron exchange are
discussed in the context of  exclusive Higgs production at the LHC.
}

\section{Introduction}
A hadronic diffractive process can be  defined as a reaction in
which no quantum numbers are exchanged between the colliding particles
and/or a large, non exponentially suppressed, rapidity gap (region devoid of
particles) is present. In the framework of  Regge theory
diffractive reactions are characterized by the exchange of a {\it
pomeron}, 
a hypothetical object with vacuum quantum  numbers.

Although the term ``diffraction'' has been used in high-energy physics
since the 1950's~\cite{history}, 
the era of exciting experimental diffractive measurements, as we know
them now, 
started in the 1980's, when ``hard diffraction'' was first
discussed~\cite{ingelman}. 
Diffractive reactions that  incorporate  hard processes, such as production of
jets in $p{\bar p}$ collisions, allow one to study diffraction in a
perturbative QCD framework, thus providing an opportunity to study 
the nature of the {\it pomeron}.

The CDF experiment at the Tevatron collider
contributed extensively to the field of diffraction during Run I
by studying  a wide range of physical processes
at two
different center of mass energies, $\sqrt{s}$=630 and 1800 GeV~\footnote{See
K. Goulianos, ``Twenty Years of Diffraction at the Tevatron,'' in these Proceedings.}.
Many important observations were made regarding the diffractive structure
function of the {\it pomeron}, the breakdown of QCD factorization in hard
diffraction between Tevatron and HERA, and the discovery of large
rapidity gaps between two jets~\cite{run1-dsf}-\cite{run1-gaps}.
In 2001, CDF started a second phase of data-taking (Run II) at $\sqrt{s}$=1960 GeV with new
upgraded detectors.

\section{CDF Forward Detectors in Run II}

\begin{figure}[h]
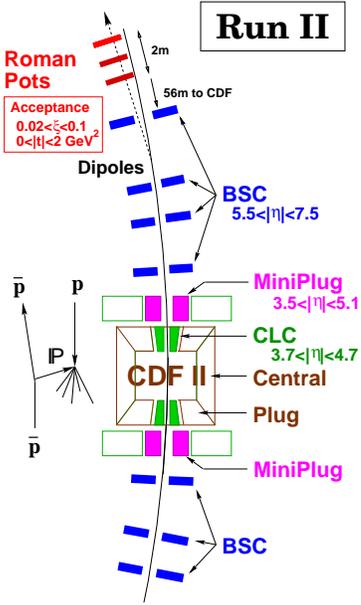

\begin{minipage}[t]{0.4\textwidth}
\protect\caption{Layout of CDF Run II forward detectors 
along the beam-pipe (not to scale).}
\label{Cdf-run2}
\end{minipage}
\begin{minipage}[b]{0.55\textwidth}
\phantom{xxx}
\end{minipage}
\end{figure}

Since the identification of diffractive events requires either
tagging of the leading particle or observation of a rapidity gap,
the forward detectors are very important for the implementation of a  
diffractive program.
The schematic layout of the CDF detectors in  Run II
is presented in Fig.~\ref{Cdf-run2}. 
\begin{minipage}[b]{0.45\textwidth}
\includegraphics[width=0.9\textwidth]{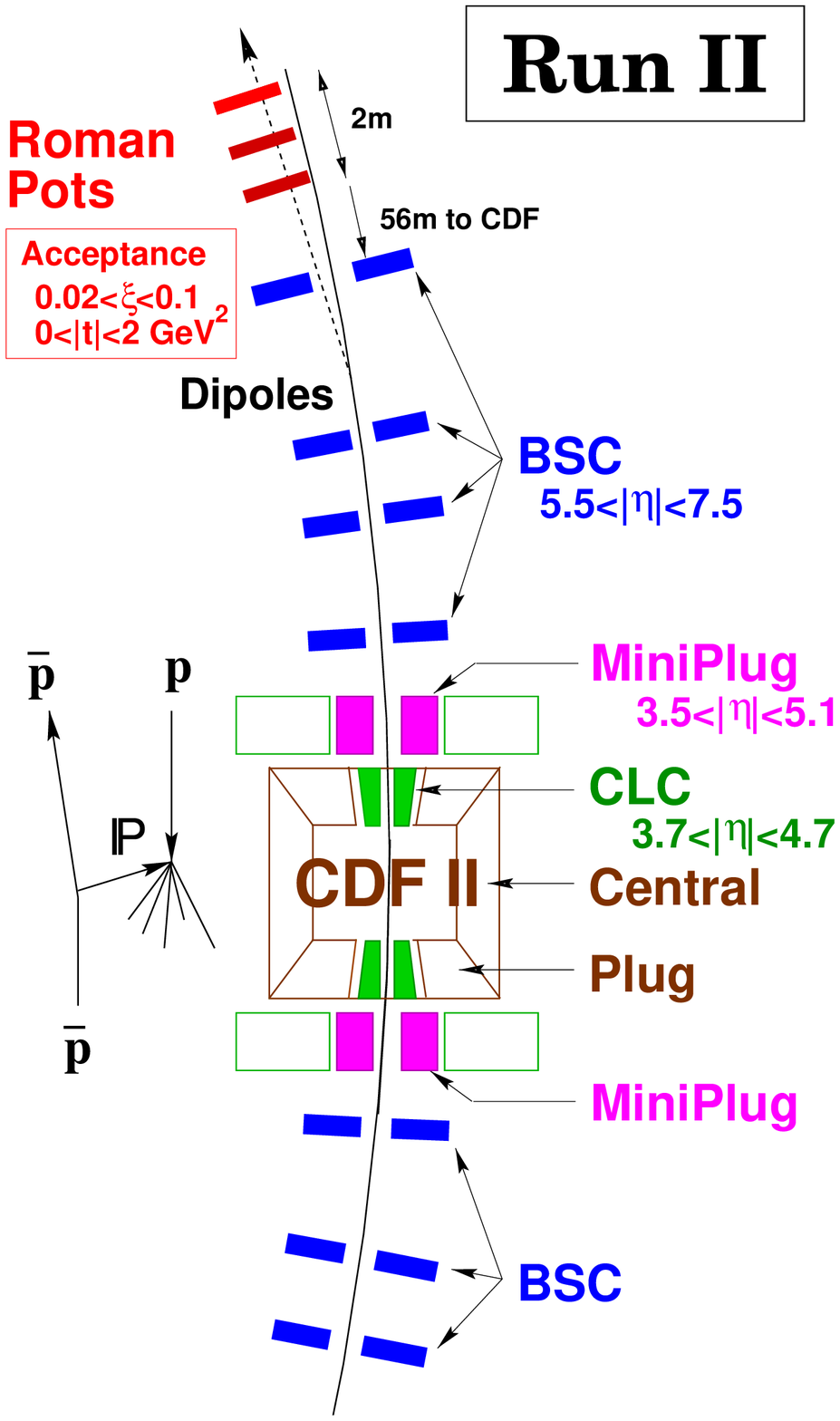}
\end{minipage}
\begin{minipage}[t]{0.55\textwidth}
\vspace{-54ex}
The Forward Detectors include the Roman
Pot fiber tracker spectrometer (RPS), the Beam Shower Counters (BSCs),
and  the Miniplug calorimeters (MP). 
The RPS was used in Run I and was re-installed for Run II to detect
leading anti-protons. It is a fiber detector spectrometer 
 located along the beamline 56~m from the interaction point (IP). It
consists
 of three stations, and a coincidence of  three trigger
counters, one in each station, selects events with a leading $\bar{p}$. 

The Beam Shower Counters~\cite{CDF-BSC}, covering the pseudorapidity range 
$5.5<|\eta|<7.5$, detect particles from the interaction point 
traveling in either
direction  along beam-pipe. There are
three(four) stations installed  at increasing distances in the
$p(\bar{p})$ directions. The BSCs  are used to
select diffractive events by identifying forward rapidity gaps and
thus reducing  non-diffractive background on the trigger level.
Two Miniplug Calorimeters~\cite{CDF-miniplug} are 
designed to measure energy and lateral
position of both electromagnetic and hadronic showers 
in the pseudorapidity region of $3.5<|\eta|<5.1$. 
The ability to measure the event energy flow in the very forward rapidity
region is extremely valuable for identification of diffractive events
in the high luminosity environment of Run II.
\end{minipage}
\vspace{-15ex}

\section{Run II Diffraction Measurements}

\subsection{Diffractive Structure Function}

 The data sample for single diffractive (SD) dijet studies is  collected by triggering on a leading anti-proton in
combination with at least one calorimeter tower with $E_T>5$ GeV.
The control non-diffractive (ND) dijet
sample is triggered on the same calorimeter tower requirement. The
ratio of SD to ND dijet production rates in leading order QCD
is equal to the ratio of the corresponding structure functions at a
given  
$x_{Bj}$, 
where  $x_{Bj}$ is evaluated for each event from the $E_T$ and $\eta$ of
the jets according to the formula: 
\(x_{Bj}=\frac{1}{\sqrt{s}}\sum_{\imath=1}^{n}
{E_T}^{\imath}e^{-\eta^{\imath}}\). 
The results are in good agreement with the Run I measurement.
Preliminary results also show that there is no significant
dependence of the ratio  on $Q^2\equiv \langle{E_T}^2\rangle$ 
in the $Q^2$ range from 100 GeV$^2$ to 1600 GeV$^2$(see Fig.~\ref{run2-rjj}(left)), which
indicates that the {\it pomeron} $Q^2$ evolution is
similar to that of the proton.

\subsection{Exclusive Dijet Production}

The possibility of  observing the Higgs boson in central exclusive production at the
Large Hadron Collider (LHC) has received  considerable attention in the
last few years (see ~\cite{Khoze}). The process
$pp\rightarrow p+H+p$, where the + sign denotes presence of a rapidity
gap, 
has many appealing properties, such as possible determination of the Higgs
mass with good accuracy and clean environment due to the presence of
rapidity gaps. Although the cross section for  exclusive Higgs
production is too small to be observed at the Tevatron, there are
several processes which could be studied to calibrate theoretical
predictions.
The two cleanest signatures are  exclusive ${\chi_c}^0$ production and
central exclusive $\gamma\gamma$ production. These processes are
under study and an upper limit of  49$\pm 18(stat)\pm39(syst)$pb
for  exclusive ${\chi_c}^0$ production from a search for
$J/\psi+\gamma$ events has been  obtained by CDF.
However, the process with the highest  rate is predicted to be  central
exclusive dijet production.

\begin{figure}[htb]
\centerline{\hbox{\psfig{figure=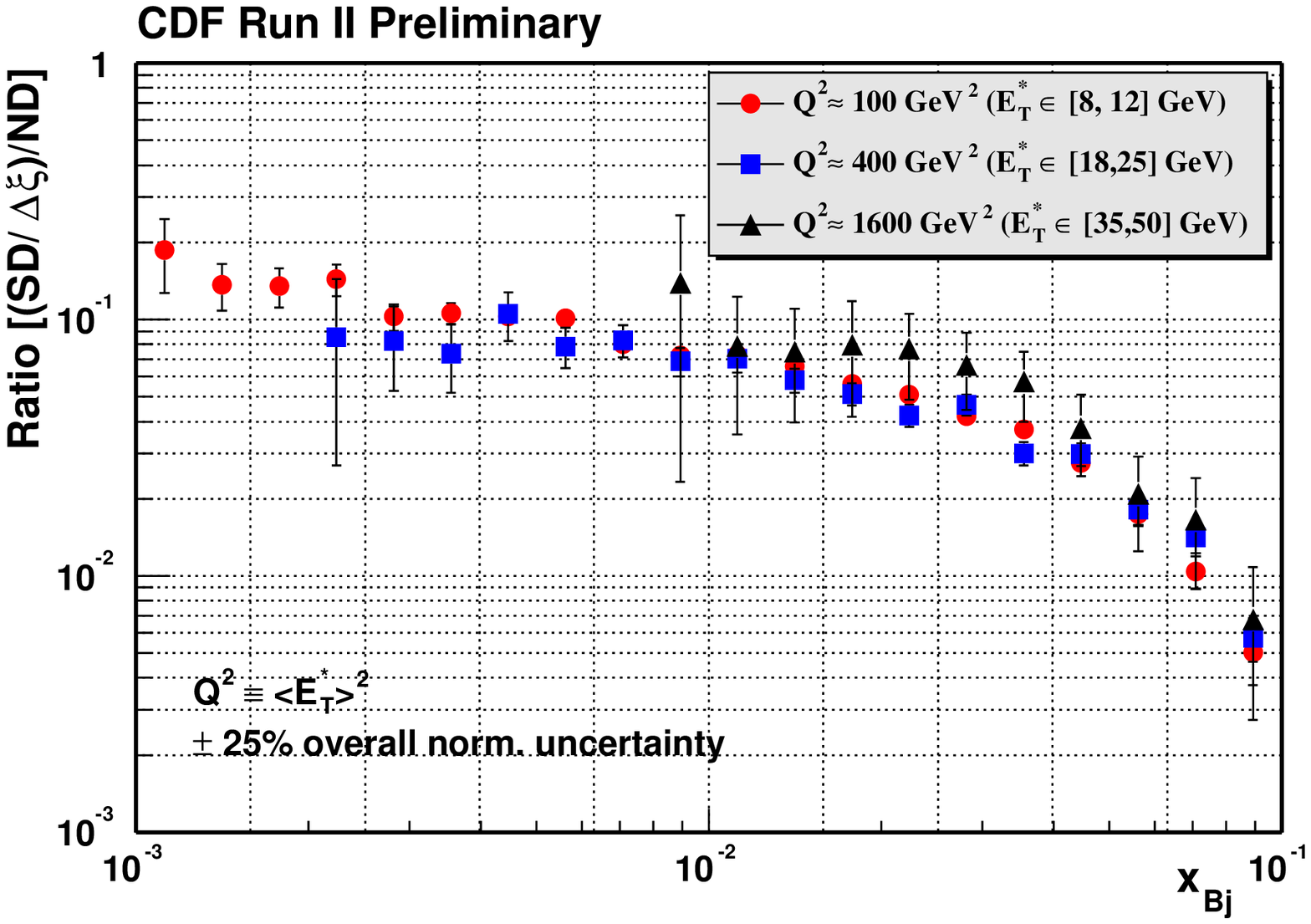,width=8cm}
                  \psfig{figure=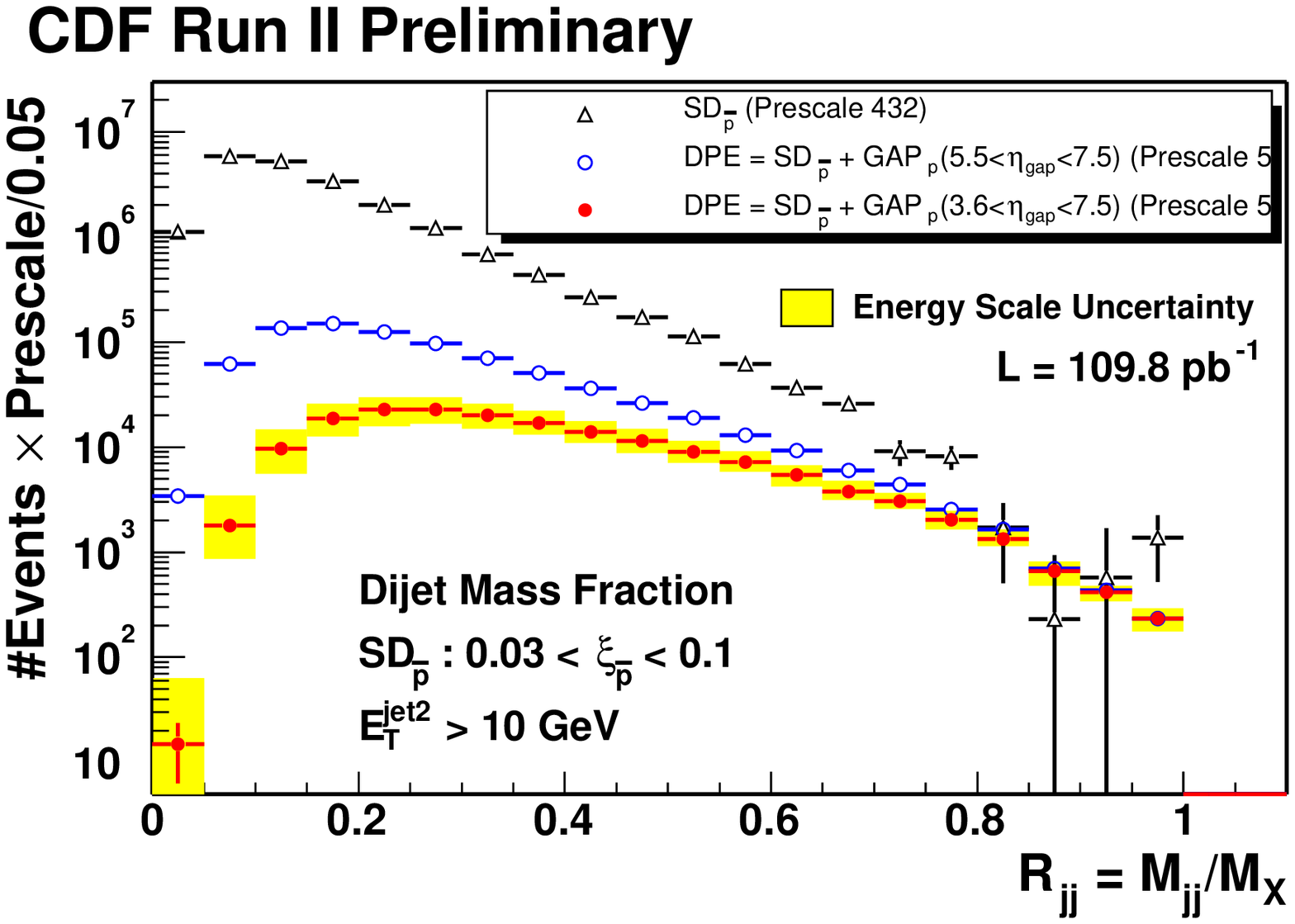,width=8cm}}}
\caption{(left) Ratio of SD to ND dijet event rates per unit $\xi$ as
a function of $x$-Bjorken at various $Q^2$ ranges; (right) Dijet mass
fraction distributions for DPE events with 
$5.5<\eta_{gap}<7.5$ (open circles) and $3.6<\eta_{gap}<7.5$ (filled
circles), and for SD events (triangles).}
\label{run2-rjj}
\end{figure}

\begin{figure}[htb]
\centerline{\hbox{\psfig{figure=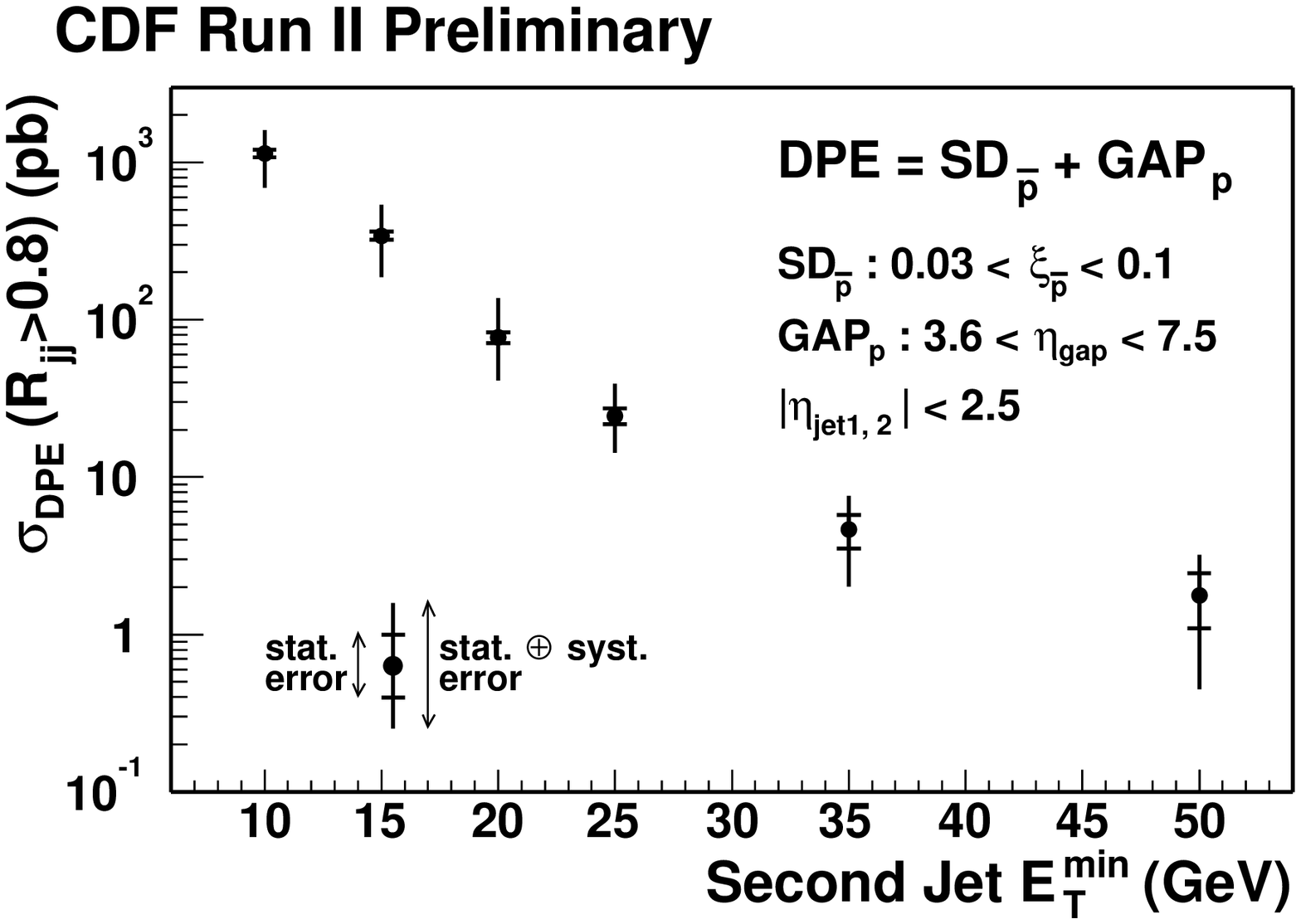,width=8cm}
                  \psfig{figure=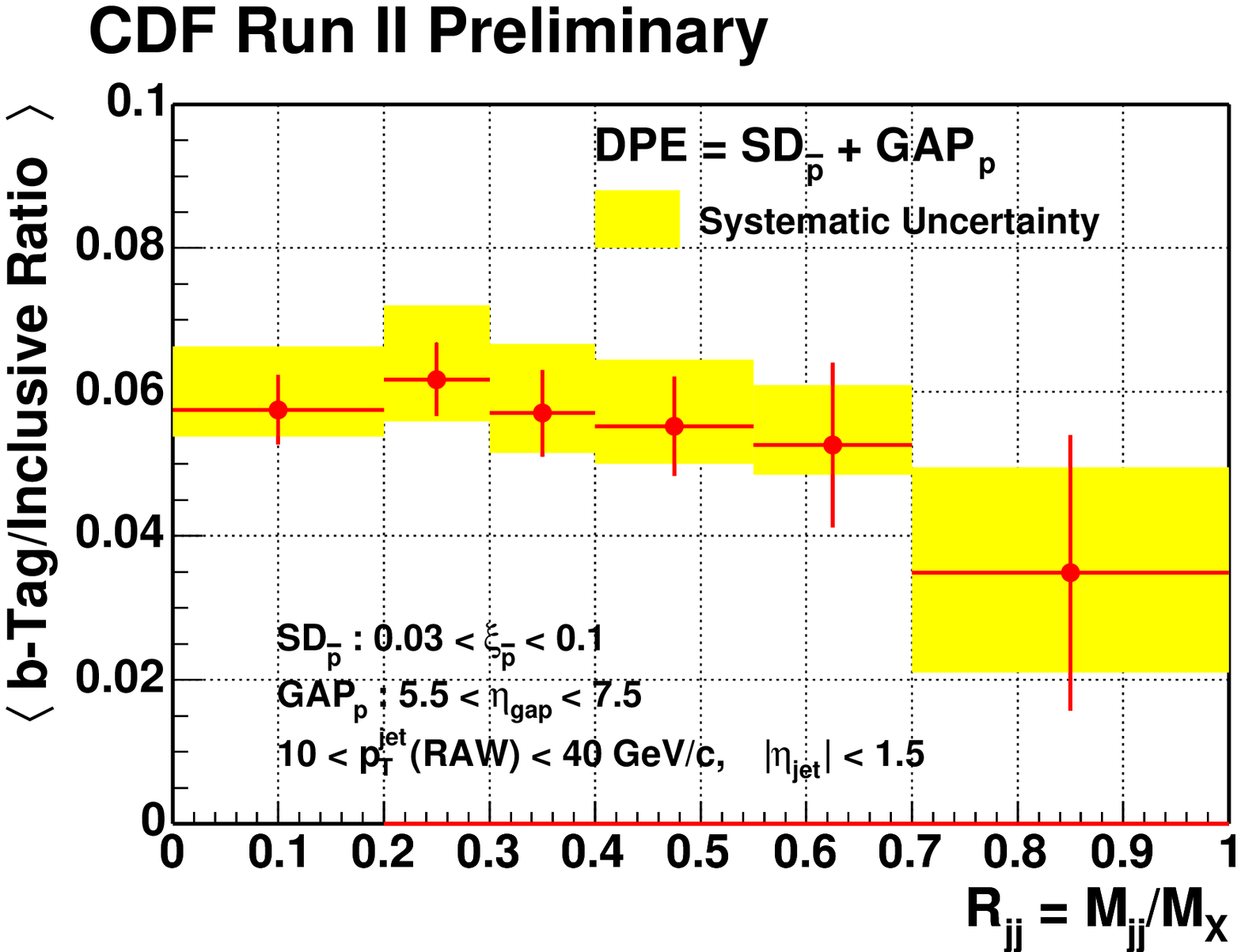,width=8cm}}}
\caption{(left) DPE dijet cross section for $R_{jj}>$0.8 as a function
of $E_T^{min}$, the $E_T$ of the next to the highest $E_T$ jet;
(right) Ratio of b-tagged jets to all inclusive jets as function of a
dijet mass fraction, $R_{jj}$.}
\label{run2-HFjets}
\end{figure}

Exclusive dijet production  by
double pomeron exchange (DPE) was studied by the 
CDF collaboration in Run I~\cite{Run1-DPE}.
In Run II our sample is collected with a dedicated trigger requiring a
BSC gap
 on the
 proton side, a leading anti-proton in the RPS, and a
single calorimeter tower with $E_T>5$ GeV. An additional offline
requirement of a gap in the MP on the proton side enhances the
signal in the
initial sample. For the resulting DPE candidate events, the ``dijet mass
fraction'', $R_{jj}$, is defined as the ratio of the invariant mass of
the two leading jets, $M_{jj}$, to the mass of the entire system, excluding
leading $p$ and $\bar{p}$, $M_X$. The dijet mass $M_{jj}$ is measured from
the energies of the calorimeter towers inside the jet cones, and the 
mass of the system is calculated according to
\(M_X=\sqrt{\xi_{p}\cdot\xi_{\bar{p}}\cdot s}\), where the values of $\xi$
are calculated from all  calorimeter towers. If the dijets were produced
exclusively, $R_{jj}$ would be, by definition, equal to unity. However,
taking into account resolution effects, the exclusive region is
experimentally 
defined as $R_{jj}>0.8$. Fig.~\ref{run2-rjj} (right) shows the $R_{jj}$ distributions
for DPE dijets, compared with those from the  single diffractive
sample. No significant excess is observed for $R_{jj}>0.8$ over a smooth
distribution.  The upper limit for exclusive dijet production, taken 
as   cross section for DPE events with $R_{jj}>0.8$, is $1.14\pm
0.06(stat)^{+0.47}_{-0.45}(syst)$ nb and $25\pm
3(stat)^{+15}_{-10}(syst)$ pb for 
leading jet $E_T>10$ and 25 GeV, respectively (see Fig.~\ref{run2-HFjets}(left)).  These experimental numbers are
consistent with recent theoretical predictions~\cite{KMR}.

One of the crucial
 advantages of  exclusive central  production is
the suppression at the leading order of 
 the background sub-process $gg\rightarrow q\bar{q}$, as
$m^2/M\rightarrow0$ 
($J_z$=0 selection rule). This condition is satisfied when the quarks
are light or when the dijet mass is much larger than the quark mass.
The $q\bar{q}$ suppression mechanism can be used to extract exclusive
dijets by identifying jets originating from quarks and looking for the
suppression of quark jets relative to all jets at high $R_{jj}$.  
CDF has 
performed this study by using jets from heavy flavor  (HF) quarks. The
advantage of the method is the relatively good efficiency of  HF
jet identification, while a disadvantage is the  necessity of
separating flavor creation HF jets from those coming from gluon splitting.
CDF studied the production of bottom quark jets in a 110 pb$^{-1}$ data
sample of DPE jet events and found  the fractions of b-tagged jets in
the inclusive
DPE dijet events, $R_{btag}$, as a function of $R_{jj}$ presented in
Fig.~\ref{run2-HFjets}(right).
A decreasing trend is observed in the $R_{jj}>$0.7 region, although no
definite conclusion can be made due to the still large statistical and
systematic uncertainties.  To quantify the magnitude of the difference
at the last bin, the ratio of the 
 weighted averages  of $R_{btag}$ for $R_{jj}>0.7$
and $R_{btag}$ for $R_{jj}>0.4$ is measured to be 0.59$\pm
0.33(stat)\pm 0.23(syst)$.
\section{Conclusion}

The CDF collaboration continues an extensive program of diffractive studies. 
The data obtained with dedicated triggers are being used to further
our knowledge of the diffractive structure function, in particular,
its $\xi$
and $Q^2$ dependence. Understanding  the 
exclusive production in DPE processes is another goal for  Run
II. Upper limits on exclusive dijet production cross sections
are comparable with  theoretical calculations. The results from 
the Tevatron data and from  interpretations of these measurements can be
applied to calibrate predictions for  central exclusive
production cross sections at the LHC.

\end{document}